\documentclass{elsart}

\usepackage{amssymb}
\usepackage{latexsym}
\usepackage{graphicx}
\usepackage{dcolumn}
\usepackage{bm}
\usepackage{setspace}
\usepackage{graphics}
\usepackage{mathrsfs}
\usepackage{amsmath}

\newcommand{\be}{\begin{equation}}
\newcommand{\ee}{\end{equation}}
\newcommand{\bea}{\begin{eqnarray}}
\newcommand{\eea}{\end{eqnarray}}

\begin{document}

\begin{frontmatter}

\title{Non-monotonic swelling of a macroion due to correlation-induced charge inversion}

\author{Brian Skinner\corauthref{cor1}},
\ead{bskinner@physics.umn.edu}
\corauth[cor1]{Corresponding author. Tel.:+1 612-624-6305; Fax: +1 612-624-4578}
\author{B. I. Shklovskii}

\address{Theoretical Physics Institute, University of Minnesota,
Minneapolis, Minnesota 55455}

\begin{abstract}

It is known that a large, charged body immersed in a solution of multivalent
counterions may undergo charge inversion as the counterions adsorb
to its surface.  We use the theory of charge inversion to examine the
case of a deformable, porous macroion which may adsorb multivalent ions
into its bulk to form a three-dimensional strongly-correlated
liquid.  This adsorption may lead to non-monotonic changes
in the size of the macroion as multivalent ions are added to the
solution.  The macroion first shrinks as its bare charge is screened and then
reswells as the adsorbed ions invert the sign of the
net charge.  We derive a value for the outward pressure
experienced by such a macroion as a function of the ion concentration in solution.
We find that for small deviations in the concentration of multivalent ions
away from the neutral point (where the net charge of the body
is zero), the swollen size grows parabolically with the logarithm
of the ratio of multivalent ion concentration to the concentration at the neutral point.

\end{abstract}


\end{frontmatter}

\section{Introduction}

The chemical and biological function of a macromolecule, and of a
polyelectrolyte in particular, often depends critically on its
ability to be stretched or compressed.  The full genomic DNA of
\textit{E. Coli}, for example, must be compacted from its contour
length of 1.4 mm (4.2 million base pairs) to fit inside a nucleolar
region about $1 \mu $m across \cite{Bloomfield}.  Achieving this
kind of compaction often requires biological organisms to expend a
great deal of metabolic energy, but in some cases a macroion may be
induced to spontaneously collapse into a tightly-packed condensate
solely through the addition of multivalent counterions to the
surrounding solution.  The trivalent ion spermidine, for example,
has been known for several decades to induce the formation of DNA
condensates \cite{Gosule}.  A number of similar examples can be
readily found for polyelectrolytes in a variety of contexts: the
collapse of polyacid brushes \cite{Li, Li-APS, Konradi} and the
formation of DNA precipitates \cite{Raspaud}, to name a few.  In
each of these cases, swelling of the polyelectrolyte can be induced
or suppressed by adding/removing ions from the surrounding solution,
thereby altering the polyelectrolyte's function.  These last two
examples, polymer brushes and DNA precipitates, provide particularly
interesting behavior, and they will be discussed more fully over the
course of this introduction.

One mechanism by which a macroion may be induced to swell or to
shrink is through correlation-induced adsorption of multivalent
ions. It is known that when a large, charged body is immersed in a
solution of multivalent oppositely-charged ions, these ions can
adsorb to its surface and form a strongly correlated liquid
\cite{overcharge}. Such correlation-induced adsorption leads to
screening of the bare charge which is much stronger than predicted
by conventional Poisson-Boltzmann solutions.  For a certain critical
value of the counterion concentration, the bare charge of the body
is totally neutralized by adsorption, and for even larger
concentrations additional adsorption inverts the sign of the net
charge.  This effect, dubbed charge inversion, has been described
extensively over the past decade (a review can be found in
Ref$.$ \cite{review}), and the theoretical framework that has been
developed is used here to describe the swelling of a charged macroion
in ionic solution.

In particular, we extend the idea of charge inversion to the case of a large, porous, highly-charged molecule---like a coiled or globular polyelectrolyte---where the ``bare charge" exists not just on the surface but within the bulk of the molecular ``ball".  In this case
ions in solution can penetrate into the volume and adsorb onto some
particular segment of the molecule \cite{WCmodel}.  If the bare charges within the ball are sufficiently dense and a sufficiently high concentration of
multivalent ions is present in the solution, adsorbed ions form a three-dimensional strongly-correlated liquid within the volume of
the ball and cause it to undergo charge inversion. Thus far the
theory of charge inversion has primarily been used to describe
surface phenomena; it our intention to apply it to a situation like
the one described here, where counterions are adsorbed within the
volume of a charged body and cause it to swell or shrink.

A number of recent experiments have observed swelling in systems of
condensed polyelectrolytes.  The authors of Refs$.$ \cite{Li} and
\cite{Konradi} have studied swelling in polyacid brushes as metal
cations are added to the surrounding solution. When salts with
divalent or trivalent cations were added, like Ca$^{2+}$ or Al$^{3+}$, the brush height was seen to decrease by as much as a factor of ten. This behavior is qualitatively similar to what is known to happen with the addition
of monovalent salt. However, the studies of Refs$.$ \cite{Li, Li-APS} and
\cite{Konradi} report non-monotonic changes in the brush height as
certain multivalent salts were added. The brush height was seen to
collapse abruptly as multivalent salt was added, reach some
minimum condensed size, and then swell again with the addition of
more salt. The non-monotonic change of height may be attributable to correlation-induced charge inversion. At some concentration of multivalent ions, the polyacid brush is neutralized and it attains its minimum size.  If
the ion concentration deviates from this neutral point, however, the
brush is either over- or under-charged and it experiences an outward
(expansive) pressure resulting from electrostatic forces.

Similar behavior was observed in studies of the structure of
DNA precipitates as the multivalent salts spermine (4+) and
spermidine (3+) were added. The authors of Ref$.$ \cite{Raspaud} performed
measurements of the interhelical spacing in liquid crystalline DNA
precipitates induced by spermine and spermidine.  In their study, short segments of nucleosomal DNA ($\sim 146$ bp) were placed in ionic solution, and at some critical concentration of multivalent salt, the DNA segments were observed to aggregate into tightly-packed condensates (bundles) of parallel, hexagonally-arranged DNA ``rods".  This arrangement is depicted schematically in the inset of Fig$.$ \ref{fig:parabola}.  As spermine or spermidine salt was added to the solution, the interaxial spacing between segments in the precipitate bundles was seen to achieve a minimum ($\sim$ 28 nm, greater than the ``hard core" diameter of double-helical DNA), and then increase again as more salt was added.  Such behavior stands in contrast to the monotonic decrease in size observed to result from the addition of monovalent NaCl ions. We suggest that in this experiment the multivalent ions form a strongly-correlated liquid within the bulk of the condensed DNA bundles, eventually leading to charge inversion and non-monotonic swelling behavior as the ion concentration is increased past the neutral point.

In this paper we consider correlation-induced adsorption in systems
of porous macroions, and we suggest a simple description of
non-monotonic swelling behavior for macroions in the presence of
multivalent salts.  We propose that at some concentration of
multivalent counterions, a porous, charged macroion achieves a
minimum size resulting from its bare charge being neutralized. As
the ion concentration deviates from this neutral point, an outward
pressure results from electrostatic forces.  The way in which this
pressure, and therefore the swelling of the polyelectrolyte, depends
on the properties of the solution is not well understood.  It is the
purpose of this paper to explore overcharge-induced swelling via a
simple model of porous charged bodies in solution, and to determine
the functional dependence of swollen size on multivalent ion
concentration near the neutral point.

The following section details the physical model to be studied and presents a full solution for the pressure and swollen size based on a free energy description of the system.  The Discussion section compares our theoretical predictions to existing experimental data and reviews the limitations of our model, followed by some concluding remarks.

\section{A simple model of swelling}

\subsection{Model and approach}

\begin{figure}
\centering
\includegraphics[width=0.50\textwidth]{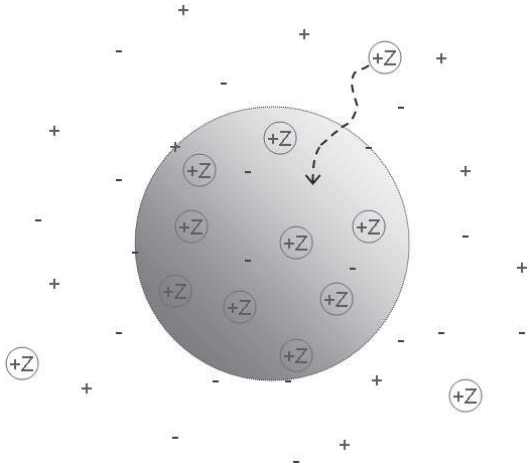}
\caption{Schematic of the system to be studied. A porous, spherical,
negatively charged body sits in a solution of positive multivalent
ions (denoted by +Z's) and dissolved monovalent salt (+'s and -'s).
Monovalent salt passes freely through the sphere while the Z-ions
may adsorb to its volume, forming a strongly-correlated liquid.}
\label{fig:schematic}
\end{figure}

We begin by considering a three-dimensional body in a solution of
multivalent counterions and some monovalent salt, such as NaCl. We
assume the body to be a sphere of radius $R$ with some fixed charge
$Q_0$, which for the sake of argument we will take to be negative,
distributed evenly throughout its volume. The body is considered to
be porous, so that both multivalent and monovalent ions can reach
and pass through its bulk. Finally, we assume that the solution
occupies a volume much larger than that of the sphere, so that any
perturbations in the electric potential or ion concentration caused
by the sphere go to zero at large distances.  A schematic of the
model is given in Fig$.$ \ref{fig:schematic}.  The choice of a
spherical geometry is arbitrary, but we note that for a cylindrical
charged body the form of the solution is the same.  The cylindrical case will be discussed more completely at the end of the following subsection.

It has been shown through numerical experiments \cite{numerics} and
physical arguments \cite{arguments} that monovalent ions respond linearly to changes in electric potential and therefore do not take
part in the formation of strongly-correlated liquids that produces
charge inversion.  On the other hand, the multivalent counterions (of valency +Z, hereafter ``Z-ions") may become strongly correlated to each other as they adsorb into the volume of the sphere.  We assume that for small deviations from the neutral point, adsorbed Z-ions are bound to the sphere by the correlation energy $\mu$, which we take to be constant for a given bare charge
density and ion valency $Z$. The limitations of this assumption are
examined in the Discussion section, but for small deviations of the
counterion concentration from the neutral point our assumption is
justified, as the properties of the strongly correlated liquid do
not change significantly.

Finally, we assume that the concentration of monovalents and
multivalent salts are fixed and constant far from the sphere.
Motivated by biological situations, we will consider the case where
the monovalent salt is much more prevalent than multivalent salt,
i.e. the monovalent salt concentration $N_{1,\infty}$ is much
greater than the Z-ion concentration $N_{Z,\infty}$.  This
assumption leads to the simplification that the Debye-H\"{u}ckel
screening radius
\be r_s = \sqrt{\frac{k_BT\epsilon}{2e^2N}} \ee
is determined solely by the concentration of monovalent salt, $N =
N_{1,\infty}$.  Here, $\epsilon \approx 80\epsilon_0$ is the
permittivity of water, $k_BT$ is the thermal energy, and
$e$ is the proton charge.  In biological situations, where salt
concentration is typically on the order of 100 mM, the
screening radius $r_s$ is less than 1 nm.  So our analysis
assumes that $r_s \ll R$.

\subsection{Free energy description}

When the sphere has a net charge, there is an expansive
pressure acting on its surface, and this pressure can be computed by
first putting together a full description of the Helmholtz free
energy $F$ of the system. Specifically, the pressure $P$ at a given
Z-ion concentration is
\be P = - \left(\frac{\partial F}{\partial V}\right)_{N_{Z,\infty}},
\label{eq:pressure} \ee
where $V \propto R^3$ is the volume of the sphere.  We are
interested in deviations in Z-ion concentration away from the
neutral point, defined as the concentration $N_Z^{(0)}$ at which the net
charge on the sphere is zero.   It is therefore natural to define
this neutral point as our reference for free energy.

In order to determine the free energy, it is necessary to solve for the electric potential in the vicinity of the sphere.  We begin by assuming that a certain net charge $Q = Q_0 + MZe$ exists within the sphere's volume, where $Q_0$ is the (negative) bare charge and $M$ is the number of adsorbed Z-ions. In actuality, the existence of a nonzero net charge represents a deviation in the number of adsorbed Z-ions from the neutral point $M_0=|Q_0|/Ze$, and the free energy associated with this deviation will be addressed momentarily. For now, however, we assume only that the sphere has some net charge $Q$, and we use it to derive the electric potential.

In the limit of large screening, $r_s \ll R$, the potential $\phi$ relative to infinity is nearly constant within the volume of the sphere and can be approximated as \cite{capacitance}
\be
\phi = \int \frac{\rho e^{-r/r_s}}{4 \pi \epsilon r} dV 
= \frac{3 Q}{4 \pi \epsilon} \frac{ r_s^2}{R^3},
\label{eq:P} \ee
where $\rho$ is the volume density of the net charge, $\rho = Q / \frac{4}{3} \pi R^3$.  This result allows us to define a an effective capacitance of the sphere 
\be C \simeq \frac{4 \pi \epsilon}{3} \frac{R^3}{r_s^2}. \label{eq:capac} \ee
If $e \phi / k_B T \ll 1$ so that the linearized Poisson-Boltzmann equation may be used, the free energy $F_{sol}$ of the ionic solution is well-known\cite{Verwey,colloidal}:
\be F_{sol} = \frac{1}{2} Q \phi . \ee

Until now we have ignored the free energy associated with the Z-ions adsorbing to the sphere, taking for granted some net fixed charge $Q$.  We now consider the process of adsorption/desorption that leads to the net charge. There is a change in both energy and entropy corresponding to the movement of Z-ions from solution to their adsorbed state within the sphere's volume.  Let us therefore ascribe some free energy $F_{ads}$ to this process so that the total free energy
\be F = F_{sol} + F_{ads} . \ee

A charge $Q$ indicates that a total number $M-M_0 = Q/Ze$ Z-ions
have adsorbed (or desorbed) to the sphere relative to the neutral
point. If this number is small compared to the total number adsorbed
onto the sphere, then we can consider the correlation energy $\mu$
and bulk counterion density $N_{Z, bulk}$ to be constant.  The total
free energy of adsorption is then
\be F_{ads} = -\frac{Q}{Ze} |\mu| + \frac{Q}{Ze} k_BT \ln\left( N_{Z,bulk} / N_{Z, \infty} \right) ,
\label{eq:Fads} \ee
where the first term on the right-hand side represents the energy
gain of the Z-ions and the second term indicates the entropic
contribution resulting from Z-ions being confined within the sphere
at a density $N_{Z,bulk}$. Eq$.$ (\ref{eq:Fads}) can be rearranged
to read
\be F_{ads} = -Q \frac{k_BT}{Ze} \ln \left( \frac{N_{Z,bulk}
\exp \left(-|\mu|/k_BT \right) }{N_{Z,\infty}} \right) .
\label{eq:Fads2} \ee
Previous authors \cite{overcharge} have noted that the quantity
$N_{Z,bulk} \exp(-|\mu|/k_BT) \equiv N_{Z,0}$ constitutes a
new boundary condition for the Poisson-Boltzmann equation in
problems where multivalent ions condense onto a charged surface.
$N_{Z,0}$ is a sort of ``vapor concentration", giving the value of
counterion concentration just above a charged surface with
correlation energy $\mu$.  In our case, this value can be used to
define a balance between $N_{Z,\infty}$ and the concentration of
adsorbed Z-ions using the Boltzmann distribution:
\be N_{Z,0} = N_{Z,\infty} \exp \left( -Z e \phi / k_BT
\right)
\label{eq:NZ0} \ee
or
\be \phi = \frac{k_BT}{Z e} \ln \left( \frac{N_{Z,\infty}}{N_{Z,0}} \right).
\label{eq:Zbalance} \ee

Eq$.$ (\ref{eq:Zbalance}) suggests that the electric potential
inside the sphere is a direct result of the concentration gradient
between the Z-ions in solution $N_{Z,\infty}$ and the ``vapor
concentration" $N_{Z,0}$. The neutral point $N_Z^{(0)}$ must
therefore occur when $N_{Z,\infty} = N_{Z,0}$.  This result allows
us to conclude that the electric potential and the amount of
overcharge (by Eq$.$ (\ref{eq:P})) are driven by the ratio of the
Z-ion concentration to the neutral point concentration.

Substituting Eq$.$ (\ref{eq:NZ0}) and (\ref{eq:Zbalance}) into Eq$.$
(\ref{eq:Fads2}), we obtain the total free energy from adsorption as
simply
\be F_{ads} = -Q \phi , \ee
so that the free energy of the entire system is just
\be F = F_{sol} + F_{ads} = -\frac{1}{2} Q \phi
\label{eq:Fcapacitor} \ee
or
\be F = - \frac{2 \pi \epsilon}{3} \frac{R^3}{r_s^2} \phi^2 .
\label{eq:Ftotal} \ee
It is worth noting that Eq$.$ (\ref{eq:Fcapacitor}) is identical in form to the free energy of a capacitor held at constant potential by a
voltage source.  In this way our sphere is like a capacitor (as in Eq$.$ (\ref{eq:capac}) ) held at constant potential $\phi$ by a concentration ratio $N_{Z,\infty}/N_Z^{(0)}$ of Z-ions.

Using Eq$.$ (\ref{eq:pressure}), we can solve for the pressure and
conclude that
\be P = - \left(\frac{\partial F}{\partial V}\right)_{N_{Z,\infty}} \\
 = - \left(\frac{\partial F}{\partial V}\right)_{\phi} = \frac{\epsilon}{2} \phi^2 / r_s^2 .
\label{eq:P2} \ee
In other words, the electrostatic potential of the sphere remains
constant as it expands, leading to a positive (expansive) pressure
that is proportional to $\phi^2$.  In terms of counterion
concentration,
\be P \propto \left[ \ln \left( \frac{N_{Z,\infty}}{N_Z^{(0)}}
\right) \right]^2 , \label{eq:PNZ} \ee
so the sphere experiences a parabolically-increasing expansive
pressure whenever the Z-ion concentration deviates from the neutral
point. We note that the pressure in Eq$.$ (\ref{eq:P2}) can be interpreted as
the energy density of the electric field at the surface of the
sphere.

We now briefly consider the case where, rather than assuming a
spherical geometry, we take the charged body to be a cylinder of
length $L$ whose radius $a \ll L$ is free to expand as multivalent
counterions bind to its volume.  We can solve for the potential using the method of Eq$.$ (\ref{eq:P}) to get $\phi_{cyl} = Q / \epsilon \pi L \cdot r_s^2/a^2$.  The pressure is therefore
\be P = - \left(\frac{\partial F}{\partial V}\right)_{\phi} = -
\left(\frac{\partial F}{\partial a}\right)_{\phi} \cdot \frac{1}{2
\pi a L} = \frac{\epsilon}{2} \phi^2 / r_s^2 . \label{eq:P2cyl} \ee
That is, the outward pressure for a cylinder has the same form as for a sphere.
This result is not particularly surprising, since for small $r_s$ the potential is constant throughout the body and therefore the net charge $Q$ and the monovalent salt are distributed isotropically throughout its interior.  In this sense the solution of charges within the body behaves like a Pascal fluid, and the outward pressure it creates (Eqs$.$ (\ref{eq:P2}) and (\ref{eq:P2cyl})) should be uniform and independent of the shape of the object containing it.

\subsection{Elastic expansion}

If some degree of elasticity is ascribed to the sphere (or cylinder), then the
outward pressure given by Eq$.$ (\ref{eq:PNZ}) causes it to
swell.  For the DNA bundles described in the Introduction, we may expect elastic behavior as the result of a balance between short-range forces within the bundle: correlation-induced attraction and steric repulsion.  When the bundle is completely neutral, its equilibrium interhelical spacing corresponds to a minimum in the total free energy.  We may expand the free energy by Taylor series to second order in the interhelical spacing and arrive at a potential that grows parabolically as the spacing deviates from this minimum (its value at the neutral point).  When the bundle becomes under- or over- charged by adsorption/desorption of counterions, the existence of a net charge creates an electrostatic pressure which produces radial swelling in this parabolic elastic potential.  For polyelectrolyte brushes, it is the brush height that swells in an elastic potential.  When the brush is neutralized, it achieves a minimum height that corresponds to a minimum of the total free energy (which includes electrostatic energy as well as conformational entropy).  The free energy near this minimum may also be expanded by Taylor series to second order in the brush height, resulting in a parabolic potential.

Our assumption of a parabolic potential for small deviations from the neutral point allows us to write down the free energy $F_B$ associated with the compressibility of the body as
\be F_B = \frac{1}{2} B x^2 ,
\label{eq:Hooke} \ee
where $B$ is the bulk modulus and $x$ is the fractional change in volume, $x = \Delta V / V_0$.  If we add this term to the free energy for our sphere we can calculate its swollen size using the
equilibrium condition
\be \left( \frac{\partial F}{\partial R} \right)_{\phi} = 0 .
\label{eq:swellingF} \ee
Let the swollen radius $R$ be written as $R = R_0 + \Delta R$, where
$R_0$ is the radius at the neutral point.  If $\Delta R$ is small
compared to $R$, so that the change in volume $\Delta V \propto
\Delta R$, Eq$.$ (\ref{eq:swellingF}) can be evaluated as
\be \frac{\Delta R}{R_0} = \frac{V_0}{3 B} \frac{\epsilon}{2}
\phi^2/r_s^2 \label{eq:swellingR} . \ee
In other words, the swollen radius is directly proportional to the
outward pressure (Eqs$.$ (\ref{eq:P2}) and (\ref{eq:PNZ}) ):
\be \Delta R \propto \left[ \ln \left( \frac{N_{Z,\infty}}{N_Z^{(0)}}
\right) \right]^2  \label{eq:swellingR2} . \ee

\section{Discussion}

\subsection{Comparison with existing data}

\begin{figure}[htb]
\centering
\includegraphics[width=0.60\textwidth]{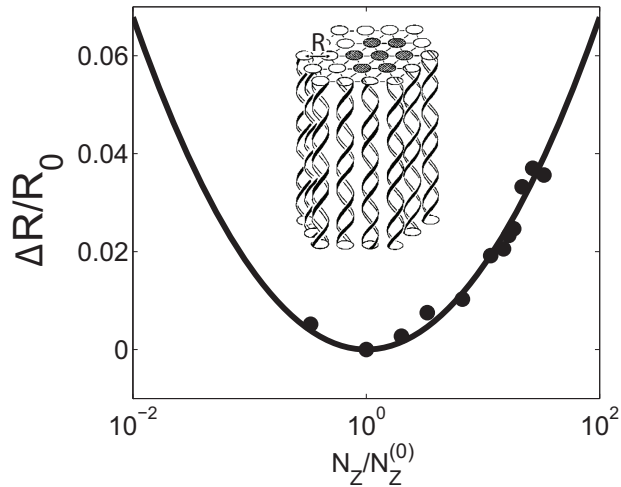}
\caption{{Swollen size of a charged body in a solution of multivalent counterions. Experimental data, shown as solid dots, is taken from
Ref$.$ \cite{Raspaud}, where adsorption of spermine salt (4+)
produces a non-monotonic change in the interhelical spacing of
condensed DNA aggregates (shown schematically in the inset, reprinted with permission from \cite{Raspaud}). The value of $N_Z^{(0)}$ is taken from Ref$.$
\cite{condensation}.  The elastic constant as defined in Eq$.$
(\ref{eq:Hooke}) is chosen to match the experimental data, and it is
similar to the elastic constant predicted by Ref$.$ \cite{Moreira}.
}}\label{fig:parabola}
\end{figure}

Fig$.$ \ref{fig:parabola} shows a typical plot of swollen radius
versus counterion concentration.  Overlaid on the plot are data from
an experiment in Ref$.$ \cite{Raspaud}, where the interhelical spacing of
closely-packed DNA condensates was measured via x-ray diffraction
for various concentrations of spermine salt (valency 4+).  In this
experiment, the addition of spermine to a solution of short DNA fragments
was seen to induce the formation of dense, ordered DNA condensates, as depicted in the inset of Fig$.$ \ref{fig:parabola}.  The interaxial spacing $a$ between parallel DNA ``rods", which in this experiment constitutes the swollen dimension, was extracted via x-ray diffraction over a range of spermine salt concentration.  The data is plotted in Fig$.$ \ref{fig:parabola} as the ratio $a/a^{(0)}$, where $a^{(0)}$ is the observed minimum spacing.

The value of the neutralizing concentration $N_Z^{(0)}$ reported in
this study ($\simeq 3$ mM) is in strikingly good agreement with the
predicted value for DNA-spermine condensates from Ref$.$
\cite{condensation} (3.2 mM), where reentrant condensation of DNA
was described in the context of correlations between adsorbed
multivalent cations.  The ``elastic constant" of the parabola in
Fig$.$ \ref{fig:parabola} was determined by fitting the experimental
data to the form of Eq$.$ (\ref{eq:swellingR}). The value of this
constant compares favorably with other theories of
correlation-induced attraction, as described in the
following section.

Data for swelling of polyelectrolyte brushes, taken from Refs$.$
\cite{Li} and \cite{Konradi} also shows non-monotonic dependence on
salt conentration. Both studies reported measurements of the height
of long polyacid brushes as divalent salts like Ca$^{2+}$ and
Mg$^{2+}$ were added. The brushes were seen to collapse abruptly
when low levels of the salts were added, and then to swell
noticeably over a certain range of higher salt concentration. There
is some disagreement between the two experiments on the degree of
swelling in the brushes, and the ability of divalent ions to form a
strongly-correlated liquid is marginal, but the convincingly
non-monotonic dependence on ion concentration is encouraging. The
authors of Ref$.$ \cite{Li} also report measurements of the brush height as
Al$^{3+}$ ions are added to solution.  They find that, over the
studied range of aluminum salt concentration, the brush is in a
collapsed state but swells noticeably with increased amounts of
Al$^{3+}$. Because of the exponential dependence of $N_Z^{(0)}$ on
$Z$, we suspect that this range (0.1 to 100 mM of AlCl$_3$) falls
above the neutral point, so that the net charge of the brush is
already inverted and further addition of trivalent salt produces
outward swelling.

\subsection{Limitations of elastic expansion}

In deriving the parabolic form of Eq$.$ (\ref{eq:swellingR2}), we
assumed that the properties of the adsorbed strongly-correlated
liquid were unchanged as the sphere expanded. The correlation energy
$\mu$ was held constant, and the sphere was assumed to respond
elastically as the radius increased. In reality, increasing the size
of the sphere causes the adsorbed ions to become more distant,
lowering the correlation energy between them.  This effect leads to
significant deviations from parabolic behavior when the Z-ion
concentration is far from the neutral point.

In the case of coiled DNA or other polyelectrolytes, the correlation energy
is in fact what holds the polymer in its condensed phase.  That is, our assumed ``elastic force" in DNA is the result of a balance between steric repulsion and correlation-induced attraction among interior charges, and therefore depends directly on $\mu$.  Thus as the polymer swells and the correlation energy between adsorbed ions decreases, the contractile ``bulk modulus" of the condensate must also decrease.  In this case the polymer will
swell faster than predicted by the elastic model.  For many polyelectrolyte condensates there will be some degree of swelling at which the repulsive electrostatic energy becomes greater than the maximum attractive correlation energy, causing the polyelectrolyte to ``blow up".  This phenomenon results in
sharp transitions between condensed and dissolved phases for DNA in solution, as has been described theoretically \cite{condensation, Solis}.

The way in which the ``elasticity" of a charged body changes with
its size may be complicated, and in general it depends on the
physical structure of the body. However, we may gain insight into
changes in the elastic energy $F_B$, and therefore of the limitations of
parabolic swelling behavior, by considering a basic example of
correlation-induced attraction. The authors of Ref$.$ \cite{Moreira} have
assembled a theoretical description of attraction between
like-charged plates in a solution of counterions that may be
relevant to our situation. Using Monte Carlo simulations, they find
scaled numerical values for the attractive pressure between parallel
plates as a function of their separation. They demonstrate that the
pressure falls off quickly when the separation between plates grows
to the order of the average distance between adsorbed ions. This
effect is similar to what we might expect for a condensed
polyelectrolyte like DNA: when the condensate swells to a certain
characteristic size the correlation energy will drop off quickly,
leading to more rapid swelling of the polyelectrolyte and eventual
resolubilization.

We can fit the numerical solution for attractive pressure between plates\cite{Moreira} to the situation of a crystalline DNA condensate (as in Ref$.$\cite{Raspaud}), where we expect the attractive force between parallel DNA helices to decrease as they are pulled apart by swelling.  Using the data shown in Fig$.$ \ref{fig:parabola}, we can evaluate the swollen size of a DNA condensate whose elasticity follows this ``parallel plate" model and predict when it might ``blow up" due to swelling.  We scale the results for pressure between plates in Ref$.$ \cite{Moreira} (their Fig$.$ 3) so that the bulk modulus $B = -V dP/dV$ at equilibrium matches the observed bulk modulus for DNA condensates near the neutral point.  In this way we determine an equation for the bulk modulus of a DNA condensate bundle as a function of the radius, $B = B(\Delta R / R_0)$.  A numerical iteration of Eqs$.$ (\ref{eq:P}) and (\ref{eq:swellingR}) can then be performed at a given Z-ion concentration in order to find the equilibrium size.

\begin{figure}[htb]
\centering
\includegraphics[width=0.60\textwidth]{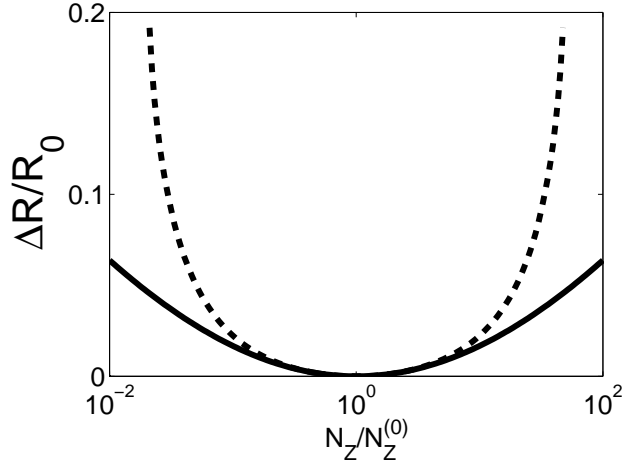}
\caption{Swelling as a function of Z-ion concentration, using an
elastic model (solid line) and a ``similarly charged plate" model
(dashed line) where elastic force decreases with swollen size. The
bulk modulus is taken from the data of Ref$.$ \cite{Raspaud} as in Fig.
\ref{fig:parabola}, and the similarly charged plate model is adapted
and scaled from Ref$.$ \cite{Moreira}.  The charged plate model deviates
significantly from the elastic model when the Z-ion concentration is
far from the neutral point, driven by a decrease in the correlation
energy between adsorbed ions.
}\label{fig:blowup}
\end{figure}

The result of this procedure is shown in Fig$.$ \ref{fig:blowup},
and highlights the sharp deviation from parabolic behavior when the
Z-ion concentration is far from the neutral point.  It is worth
noting that the bulk modulus resulting from fitting experimental data (6.5 MPa) is similar to what one might estimate \cite{estimate} based on the parallel-plate model of Ref$.$\cite{Moreira} for the equilibrium point (13 MPa). The experimental data presented in Fig$.$ \ref{fig:parabola}, however, does not deviate from the elastic model as strongly as the prediction of Fig$.$ \ref{fig:blowup}, and this difference may be attributed to the increased flexibility and conformational degrees of freedom of the DNA segments as compared to two parallel plates.  In fact, the dotted line in Fig$.$\ref{fig:blowup} represents a sort of ``worst-case scenario" for deviation from the simple parabolic description of Eq$.$ (\ref{eq:swellingR2}), where the system has no freedom to change conformation and the correlation energy must decrease abruptly with swollen size.

\section{Concluding remarks}

We have shown in this paper how a porous macroion may exhibit
non-monotonic swelling behavior as multivalent ions are added to the
surrounding solution.  We have presented a simple model of swelling
near the neutral point based on the concept of correlation-induced
adsorption of ions, and concluded that the swollen size of the
macroion grows as the square of the logarithm of the ion
concentration.  

As a final comment, we note that the applicability of our conclusion depends strongly on the ability of adsorbed multivalent counterions to form a strongly-correlated liquid.  The degree to which this is possible depends in general on the background to which the multivalent counterions can adsorb.  If the ``bare charges" distributed throughout the macromolecule are too distant, then such a strongly-correlated liquid will not be possible; the distance between adsorbed Z-ions must remain smaller than the screening radius $r_s$ or there will be no significant correlation energy.  The valency of the counterions also plays an important role: the correlation energy $\mu$ of condensed counterions grows directly \cite{overcharge} with $Z$, and the concentration of multivalent ions required to produce overcharging decreases exponentially with $\mu$.  Theoretical estimates \cite{overcharge, strongcoupling} indicate that our approach is valid for $Z \geq 3$, while the case of $Z = 2$ is marginal.  A recent paper \cite{Kundagrami} has specifically considered the case of divalent ions adsorbing to a flexible polyelectrolyte, including the effects of correlations between adsorbed ions along the polyelectrolyte backbone.  The authors conclude that even a modest concentration of divalent salt can produce overcharging and non-monotonic swelling at room temperature.

\begin{small}
\vspace*{2ex} \par \noindent
{\em Acknowledgements.}

We are grateful to M. Rubinstein, E. Zhulina, T. Nguyen, and R. Zhang for helpful discussions.

Brian Skinner acknowledges the support of the Phi Kappa Phi Yoerger
Presidential Fellowship and the Anatoly Larkin Fellowship.
\end{small}


\newpage
\listoffigures


\begin{thebibliography}{100}
\bibitem{estimate}Such an estimate can be made by assuming that parallel DNA ``cylinders" within the condensate interact like parallel plates of fixed surface charge.  Geometric corrections will tend to reduce this estimate, bringing it into better agreement with the observed data.
\end{thebibliography}
\end{document}